\documentclass{article}
\usepackage[utf8]{inputenc}

\usepackage{mathtools}
\numberwithin{equation}{section}

\usepackage{pdfsync}
\usepackage{dsfont}
\usepackage{xcolor}
\usepackage{cancel}
\usepackage{esint}
\usepackage{amsthm}
\usepackage{enumerate}
\usepackage{amsfonts}
\usepackage{amssymb}%
\usepackage{braket}
\usepackage{bm}
\usepackage{amsmath,amsthm}
\usepackage{graphicx}
\usepackage{caption}
\usepackage{subcaption}
\usepackage{enumitem}
\usepackage[bottom]{footmisc}

\usepackage{geometry}
\newgeometry{vmargin={35mm}, hmargin={35mm,35mm}}

\usepackage{hyperref}
\usepackage[backend=biber, style=numeric, sorting=nyt]{biblatex}
\usepackage{amsfonts}
\usepackage{graphicx}%
\providecommand{\U}[1]{\protect\rule{.1in}{.1in}}
\newtheorem{theorem}{Theorem}[section]
\newtheorem{proposition}{Proposition}[section]
\newtheorem{lemma}{Lemma}

\newcommand{\R}{\mathbb{R}}
\newcommand{\C}{\mathbb{C}}

\definecolor{mygreen}{rgb}{0.1,0.75,0.2}

\newcommand{\eps}{\varepsilon}

\newcommand{\pv}{\text{P.V.}}

\title{A modern approach to the Kelvin-Helmholtz instability on circular vortex sheets}
\author{
    Galen Wilcox\thanks{\normalsize\itshape
   MIT-WHOI Joint Program in Oceanography/Applied Ocean Science \& Engineering, Cambridge and Woods Hole, MA, USA} ~\& Ryan Murray\thanks{\normalsize\itshape
   North Carolina State University, Raleigh, NC, USA}}
\date{\today}

\begin{document}

\maketitle


\begin{abstract}
    We represent the outermost shear interface of a eddy by a circular vortex sheet in two dimensions, and provide a new proof of linear instability via the Birkhoff-Rott equation. Like planar vortex sheets, circular sheets are found to be susceptible to a violent short-wave instability known as the Kelvin-Helmholtz instability, with some modifications due to vortex sheet geometry. This result is in agreement with the classical derivation of \cite{Moore1974,SaffmanVortex}, but our modern approach provides greater clarity. We go on to show that the linear evolution problem can develop a singularity from analytic initial data in a time proportional to the square of the vortex sheet radius. Numerical evidence is presented that suggests this linear instability captures the wave-breaking mechanism observed in nonlinear point vortex simulations. Based on these results, we hypothesize that the Kelvin-Helmholtz instability can contribute to the development of secondary instability for eddies in two-dimensional turbulent flow.
\end{abstract}

\section{Introduction}

Sharp shear interfaces, characterized by a mismatch of tangential velocities along a surface (in three dimensions) or a curve (in two dimensions), are ubiquitous in real fluid flows. Important examples of such interfaces arise in engineering, including the wake of an airfoil, buoyant outflow of steam from a pipe, or industrial mixing processes. In nature, some examples include rocks interrupting the flow of a river, flow along density interfaces in stratified fluids, or the convergence of rivers like the Rio Negro and Rio Solim\~oes in Brazil. Shear interfaces also arise at large scales where planetary rotation is important, for  example at the margins of ocean currents like the Gulf Stream or swirling bands of gas on the planet Jupiter. Our understanding of these shear interfaces benefits from idealized mathematical analysis.

It is well-known that flat interfaces between inviscid, incompressible fluids with discontinuous tangential velocities are susceptible to the Kelvin-Helmholtz instability. This is a ``wave-breaking" instability which magnifies perturbations to such a shear interface. The interface steepens until it is no longer a function of the streamwise coordinate, and subsequently rolls up into tight spiral vortices. The Kelvin-Helmholtz instability can thus contribute to the generation of eddies which are characteristic of two-dimensional turbulent flow, for example in the separated shear layer behind an airfoil \cite{michalke1967inviscid, Yarusevych2006} or at density interfaces in oceans and rivers \cite{geyerRiversKHI,woods_1968}. However, the eddies generated by the Kelvin-Helmholtz instability do not remain organized for long, and often develop a secondary instability that results in a transition to smaller-scale turbulence \cite{geyerRiversKHI,klaassen_peltier_1985, thorpe_1973}. 

Here, we apply modern analytical vortex sheet methods to prove that as an idealized model of an eddy, circular vortex sheets are also susceptible to the Kelvin-Helmholtz instability. We therefore hypothesize that the same mechanism can both generate eddies in a two-dimensional shear flow, and lead to their breakdown when further perturbations are applied. Our approach improves upon a previous derivation of the same instability \cite{Moore1974, SaffmanVortex}. We obtain a matrix solution which describes the short-time development of instability, allows us to explicitly identify singularity formation in finite time, and agrees with nonlinear simulations up to the time of singularity development.

\subsection{The Birkhoff-Rott equation for vortex sheet evolution}

We begin with a review of some vortex sheet fundamentals. Consider the simple shear flow along a flat interface described by
\begin{equation}
u_F(x,t) \equiv u_F(x) = \begin{cases} (u^+,0) &\text{ for } x_2 > 0 \\ (u^-,0) &\text{ for } x_2 < 0 \end{cases},
\end{equation}
where the spatial coordinates are $x=(x_1,x_2)$. The subscript ``F" simply denotes ``flat." Due to the discontinuity along the surface $x_2=0$ this velocity field is not a classical solution of the incompressible Euler equation
\begin{equation}
\partial_t u + u\cdot \nabla u = \nabla p,\qquad \nabla \cdot u = 0.
\end{equation}
However, by rewriting this equation in terms of the vorticity $\omega = \nabla \times u$, we arrive at the system
\begin{equation}
\label{eqn:vorticity-stream}
    \begin{split}
        & \partial_t\omega + u \cdot \nabla \omega = 0,\\
        & u = \int K(x-y) \omega(y)dy,\\
        & K(z) := -\frac{1}{2\pi} (-z_2/|z|,z_1/|z|).
    \end{split}
\end{equation}
Equation \eqref{eqn:vorticity-stream} can be interpreted to mean that the vorticity is purely advected by the velocity field in two-dimensions, and that the velocity is determined by a non-local linear operator, namely the Biot-Savart law, acting on the vorticity. This is known as the vorticity-stream formulation of the Euler equation. The vorticity-stream formulation admits a very weak interpretation in which vorticity is permitted to be a measure supported on an injective curve, as opposed to only a function. We write such a curve in the complex plane as $z:[a,b]\times [t_0,t_1] \to  \C \sim \R^2$, parametrized by vorticity density, and call this curve a vortex sheet. In the case of the flat vortex sheet, $[a,b]=\R$.

By manipulating Equation \eqref{eqn:vorticity-stream} one arrives at the Birkhoff-Rott equation
\begin{equation}
    \partial_t z^*(\alpha,t) = \frac{1}{2\pi i}\pv \int_a^b \frac{1}{z(\alpha,t) - z(\beta,t)} \,d\beta,
    \label{e:bre}
\end{equation}
where $\pv$ denotes the principal value integral, which is necessitated by the order of the singularity in the integrand as $\beta \to \alpha$. The notation $(\cdot)^*$ represents the complex conjugate. For simplicity in this work we only consider closed or infinite vortex sheet curves. A derivation of the Birkhoff-Rott equation from the Euler equation can be found in \cite{wuBRE}; this equation provides an advantageous starting point for vortex sheet analysis because it gives direct insight into the evolution of the interface.

In particular, the steady flow $u_{F}$ corresponds to a flat vortex sheet of infinite extent, where vorticity is a measure supported on the line $x_2=0$ and density along the line proportional to $(u^--u^+)$. This sheet can be simply described in the complex plane and its stability assessed via the Birkhoff-Rott equation; details are given in Section \ref{s:flat_sheet}. For now it suffices to say that after linearizing about the steady solution and applying a small perturbation, Fourier analysis proves that a strong instability develops in finite time \cite{BM}, intensified at short wavelengths. Linear theory is thus able to predict the strong instability demonstrated by observations and numerical experiments.

Consider a two-dimensional eddy generated via the Kelvin-Helmholtz instability from a flat shear interface. As in \cite{michalke1967inviscid}, we suppose that the essential characteristics of such an eddy can be described by a circular shear layer. We further idealize the eddy as a circular vortex sheet; this vortex sheet will be the focus of our analysis. 

It is reasonable to consider geometric perturbations to the circular vortex sheet, which should lead to the development of smaller vortices via the Kelvin-Helmholtz instability in a similar manner as for the flat sheet. This repeating process of vorticity consolidation and redistribution via shear instability could serve as a mechanism for secondary instability. The eventual result is a turbulent flow containing eddies at a range of spatial scales. Such a process is permitted by the theory of two-dimensional turbulence; squared vorticity (enstrophy) may not be dissipated without viscosity, but may cascade to smaller scales when material surfaces are extended \cite{batchelor_1969}. 

The flow field defining a circular vortex sheet is 
\begin{equation}\label{e:circ_flow_field}
u_C(x,t) \equiv u_C(x) = \begin{cases} u^+ \frac{x^\perp}{|x|^2} & \text{ for } |x| > R \\ u^- \frac{x^\perp}{|x|^2} & \text{ for } |x| < R \end{cases},
\end{equation}
where $u^+,u^-$ are again scalars with $u^+ \neq u^-$. We remark that utilizing the Birkhoff-Rott formulation corresponds to the situation where $u^-=0$, as otherwise there will be an implicit point vortex at the origin. The circular vortex sheet has physical relevance, exhibiting similarity to vortex shedding in \cite{pierce_1961} and to the sharp shear layers observed in geophysical flows, such as hurricane eyewalls or mesoscale (\textit{O}(100 km)) ocean eddies on Earth \cite{zhang2015} and the polar vortices of gas giants \cite{aguiar,cassiniImages,Mitchell2021}. One particularly compelling geophysical application of our central hypothesis is the development of submesoscale (\textit{O}(20 km)) ocean eddies via the horizontal shear instability of mesoscale eddies \cite{munk2000,annurevSubMeso}. There is the important caveat that stratification and planetary rotation must not be neglected in geophysical flows; we do not include these effects in our analysis. Despite this simplification, the analogy of geophysical flow to two-dimensional turbulence is striking, and the dynamics discussed here may find qualitative expression at large scales \cite{rhines79}.

This paper presents a modern approach to the Kelvin-Helmholtz instability on circular vortex sheets, demonstrating the wave-breaking mechanism via linear analysis and providing a strong foundation for further study. Though the stability of a circular vortex sheet has been studied previously, the problem fell stagnant for some time and the older literature is limited. We discuss this literature in the following subsection. In Section \ref{s:flat_sheet}, we review the modern analytical approach to the stability of flat vortex sheets. In Section \ref{s:circular_sheet}, we present new linear analysis for circular vortex sheets which parallels that approach, prove that the circular vortex sheet is susceptible to the Kelvin-Helmholtz instability, and demonstrate finite-time singularity formation for the linear equation from analytic initial data. In Section \ref{s:numerical_demos}, we show nonlinear numerical experiments which, by comparison with the linear solution, suggest that the linear dynamics are sufficient to capture the initial wave-breaking mechanism of the Kelvin-Helmholtz instability prior to spiral roll-up.

\subsection{Related work}

The evolution of vortex sheets has been widely studied in the mathematical literature. The derivations and an in-depth account of many analytical properties of the Birkhoff-Rott equation can be found in the classical book \cite{BM}, as well as in the introduction of \cite{wu2006mathematical}.

It is well-known that the Birkhoff-Rott equation is rather delicate, and is ill-posed in the sense of Hadamaard (in appropriate $C^k$ or $H^s$ spaces) \cite{caflisch1989singular,duchon1988global}. Indeed, one can show that, for closed vortex sheets, if a solution is $C^{1+\rho}$ then it must also be $C^\infty$ \cite{lebeau2002regularite}. On the positive side, by a Cauchy-Kowalevski type argument, it is known \cite{lebeau2002regularite} that if a curve is initially analytic then there exists a solution to the Birkhoff-Rott equation for short time. Conditions establishing the equivalence between solutions of Birkhoff-Rott and a weak version of Euler's equation is given in \cite{lopes2007criterion}.  The connection between solutions of the incompressible Euler equation with vortex sheet initial data and smoothed approximations of the same have also been studied by a number of authors, see e.g. \cite{liu1995convergence,lopes2007criterion}.

The Kelvin-Helmholtz instability was first studied in the late 19th century. The modern approach and derivation from the Birkhoff-Rott equation, which has an elegant connection with Fourier analysis and Hilbert transforms, appeared in \cite{duchon1988global} and is clearly presented in Chapter 9 of \cite{BM} as well as Chapter 8 of \cite{SaffmanVortex}. Many of the early works about well- and ill-posedness centered on perturbations of the flat sheet. This was partially driven by a number of classic numerical studies which provided evidence for spiral roll up from flat sheets \cite{krasny1986study,moore1985numerical}. Experimental observation of the instability and subsequent roll-up of the flat sheet were also well-known at the time, see e.g. the classic book \cite{drazin_reid_2004}. This type of spiral roll-up is expected to be self-similar, and a wide range of numerical and analytical work has sought to better understand self-similar solutions to Euler's equation which resemble vortex sheets \cite{bressan2020self,elling2016algebraic,elling2016self,pullin1978large,pullin1981generalization}. One group of authors has recently focused on the analysis of logarithmic spiral vortex sheets by the Birkhoff-Rott equation \cite{cieslak2022existence,cieslak2024instab,cieslak2021well}, demonstrating that spiral vortex sheets are also susceptible to the Kelvin-Helmholtz instability, albeit with a much weaker algebraic growth rate.

The analysis we present here has strong connections to some older works in the fluid dynamics literature. The classical work \cite{michalke1967inviscid} demonstrates the instability of a circular shear interface with respect to velocity perturbations: that is they study the stability of solutions to Euler's equation perturbing off of the velocity profile given by a circular shear flow. The authors motivate their analysis by considering a fully-developed spiral vortex, and assume at the top of page 655 that the shape of the discontinuity sheet is constant in time. Since they consider velocity perturbations without allowing changes in geometry, vorticity is induced away from the discontinuity sheet and thus the flow can no longer be modeled as a vortex sheet. We study a rather different situation in the present work because we allow geometric perturbations to the vortex sheet and insist that vorticity remains confined to the perturbed contour.

This project builds upon the classical work of \cite{Moore1974, SaffmanVortex}. Those authors worked closely together to develop analysis for an expanding circular vortex sheet under geometric perturbations by a potential flow approach, from which the instability of a non-expanding sheet can be deduced as a special case. Using an analogous ansatz to the form we consider here, they prove that short wave perturbations are unstable in the sense of unbounded exponential growth. Unfortunately, their analysis lacks detail and utilizes a pressure condition that, by taking only the tangential derivative rather than following the perturbed vortex sheet, neglects linear order terms (Section 8.2, pg 144, Equation 13 in \cite{SaffmanVortex}). In the present work, we obtain short-wave instability by a completely different method which parallels the modern approach to the flat vortex sheet. We find the exponential growth rate to be in agreement with their result, but are able to make further deductions due to the explicit form of our solution.


Vortex sheet analysis by the Birkhoff-Rott equation holds some distinct advantages. Compared to the potential flow approach of \cite{Moore1974, SaffmanVortex}, analysis via the Birkhoff-Rott equation is compelling in its simplicity and inclusion of the salient flow conditions. In the Birkhoff-Rott framework, one can ignore the dynamics away from shear interfaces, which may be themselves unstable, and instead focus on a simpler equation which gives direct information about the evolution of the interface. The correct pressure, kinematic, and potential flow conditions are included by construction \cite{BM}. Complex analysis can also make the computations quite elegant, particularly when working with closed vortex sheet contours. Compared to directly solving the incompressible Euler equations, the Birkhoff-Rott framework has produced more explicit vortex sheet solutions, and thus led to stronger results concerning existence, uniqueness, and singularity formation \cite{wu2006mathematical}. In fact, a later work of Moore used the Birkhoff-Rott equation to prove singularity formation for the flat sheet \cite{moore_singularity}. 

Outside of the infinite flat and closed circular sheet there are relatively few explicit solutions to the Birkhoff-Rott equation. Two notable examples are limits of the rotating Kirchhoff elliptic vortex patch (described, e.g. in \cite{batchelor2000introduction}) and the Prandtl-Munk vortex sheet \cite{munk1919isoperimetrische}. These can be used as a building block for other families of vortex sheets \cite{protas2020rotating}. It was recently shown that the instability of these solutions with respect to to geometric perturbations can be rather weak, corresponding to algebraic growth \cite{protas2021finite}. Their conclusion is that, for the limit of Kirchhoff ellipses and for the Prandtl-Munk vortices that the instability grows stronger in the wavenumber, and hence is similar in character to the flat sheet Kelvin-Helmholtz instability, a phenomenon that we confirm for the circular vortex sheet. 

In Section \ref{s:numerical_demos}, we show simple numerical experiments to demonstrate the full nonlinear manifestation of the Kelvin-Helmholtz instability on a circular vortex sheet. Numerical methods for vortex sheets require delicate attention, especially in regions where the sheet nearly self-intersects: this can be immediately seen by noticing that the need for a principal value integral in the definition of the Birkhoff-Rott equation. Important classical works regarding numerical approximation of vortex sheets include \cite{krasny1990computing,nitsche2003comparison}. Often these methods boil down to discretizing the vortex sheet into point vortices and solving for their evolution with standard Runge-Kutta methods. For many classical examples of incompressible vortex sheet evolution, including the Kelvin-Helmholtz instability, it is necessary to replace the singular kernel with a regularized approximation. An excellent overview of this process is presented by \cite{krasny_2008}. 

The regularization approach also connects directly with other non-local and kernel-driven evolution equations, such as the $\alpha$-Euler equations and aggregation equations: these have also been generalized to vortex sheet evolution and studied analytically and numerically in \cite{bardos2010mathematics,sun2012generalized}. We note that forward in time analytical properties of motion driven by these regularized kernels tend to be much better than those of the standard Birkhoff-Rott equation, see, e.g. \cite{bardos2010global}. A number of computational studies have demonstrated instability reminiscent of Kelvin-Helmholtz for the circular vortex sheet in regularized simulations of the nonlinear dynamics. Recently \cite{sun2012generalized} numerically studied the (in)stability of circular vortex sheets for a variety of kernels, including the standard Birkhoff-Rott kernel as well as kernels from aggregation equations. Similar numerical experiments were carried out by \cite{Sohn2013}.

Experimental studies of the Kelvin-Helmholtz instability on circular shear layers tend to focus on geophysical flows where background rotation is important. In this situation the Kelvin-Helmholtz instability equilibriates into patterns of polygonal vortices \cite{aguiar,fruh_read}. Some of this work is inspired by the observation of a stable hexagonal structure on the north polar surface of Saturn, composed of clouds in relative motion. This structure has existed without any significant changes since it was first observed by the Voyager spacecraft in 1988 \cite{cassiniImages,godfrey}. Though these geophysical flows bear strong qualitative resemblance to our numerical experiments in Section \ref{s:numerical_demos}, their persistence is due to the Coriolis effect, which is not included in the present work. Through careful experimental design, \cite{rabaud} ensures that neither centrifugal nor Coriolis forces dominate the dynamics. In this case the underlying physics are aligned with the analysis we present here, and indeed, the Kelvin-Helmholtz instability is observed.

Our previous work \cite{RyanNonlinear} explores nonlinear instability mechanisms for circular vortex sheets. Our main result is that a quadratic term representing the complex Burgers equation can drive singularity formation in high-frequency regimes. The nonlinear study was motivated by a quadratic transport equation derived from the Birkhoff-Rott equation, working under the erroneous conclusion that the linear dynamics were stable. After publication, we identified a critical error in the derivation of the linear and nonlinear transport equations, the details of which are discussed in a short corrigendum \cite{Ryancorrigendum}. The upshot is that the nonlinear instability mechanisms we discuss therein are not directly connected to the Birkhoff-Rott equation and the problem of circular vortex sheet evolution. The complex Burgers structure may still hold relevance to the nonlinear problem, but the linear instability problem is better approached by Fourier methods. Here, we return to the linear dynamics, and present a mechanism that explains the observed instability of circular vortex sheets up to the point of wave-breaking.


\section{Flat vortex sheets}\label{s:flat_sheet}

In reviewing the linearized theory for the Birkhoff-Rott equation near flat vortex sheets our description will follow the method given in \cite{BM}. While this theory is already well-known, it offers an important starting point when we examine the circular vortex sheet in the next subsection. We begin by considering the perturbation of the flat sheet $z:(-\infty,\infty)\times [t_0,t_1] \to  \C \sim \R^2$ described by $z(\alpha,t)=\alpha+\xi(\alpha,t)$, where $\xi$ is uniformly smooth and small in $\mathbb{C}$. For $\xi\equiv0$ we have a constant solution to the classical Birkhoff-Rott Equation \eqref{e:bre}. For a nonzero perturbation we obtain
\begin{align}
    \partial_tz^*(\alpha,t)=\partial_t\xi^*(\alpha,t)&=\frac{1}{2\pi i}\pv\int_\mathbb{R}\frac{1}{\alpha+\xi(\alpha)-\beta-\xi(\beta)}d\beta\\
    &=\frac{1}{2\pi i}\pv\int_\mathbb{R}\frac{1}{\alpha-\beta}\frac{1}{1+\mathcal{E}}d\beta,
\end{align}
where
\begin{equation}\mathcal{E}=\frac{\xi(\alpha)-\xi(\beta)}{\alpha-\beta}.
\end{equation}
Since the quantity $\mathcal{E}$ approximates the difference quotient when $\beta$ approaches $\alpha$, the smoothness of $\xi$ implies that $\mathcal{E}$ is also small. Therefore, we linearize about $\mathcal{E}=0$ to obtain
\begin{equation}\frac{1}{2\pi i}\pv\int_\mathbb{R}\left( \frac{1}{\alpha-\beta} -\frac{\xi(\alpha)-\xi(\beta)}{(\alpha-\beta)^2}\right) d\beta.
\end{equation}
Using the fact that
\begin{equation}\pv\int_\mathbb{R}\frac{1}{\alpha-\beta}d\beta=0
\end{equation}
and applying integration by parts on the second term, we obtain
\begin{equation}
    \partial_t\xi^*(\alpha,t)=\frac{1}{2}\mathcal{H}\partial_\alpha\xi(\alpha,t),
    \label{e:hilbert_form_lin}
\end{equation}
where $\mathcal{H}$ is the Hilbert transform
\begin{equation}\mathcal{H}f(x)=\frac{1}{\pi i}\int_\mathbb{R}\frac{f(y)}{y-x}dy.
\end{equation}
If we write $\xi$ as the Fourier mode
\begin{equation}\xi(\alpha,t)=A_k(t)e^{ik\alpha}+B_{k}(t)e^{-ik\alpha},~k>0,
\end{equation}
then the Hilbert transform of $\partial_\alpha\xi$ is given by
\begin{equation}\mathcal{H}\partial_\alpha\xi(\alpha,t)= ikA_k(t)e^{ik\alpha}+ikB_{k}(t)e^{-ik\alpha}.
\end{equation}
For further discussion of the Hilbert transform in the context of this problem, see \cite{BM}. This transform is then used to solve the linearized Equation \eqref{e:hilbert_form_lin}, yielding expressions for the Fourier coefficients
\begin{align}
    &A_k(t)=c_1e^{kt/2} + c_2e^{-kt/2},\label{e:soln_flat_1}\\
    &B_{k}(t)=-ic_1^*e^{kt/2} + ic_2^*e^{-kt/2},\label{e:soln_flat_2}
\end{align}
where $c_1,~c_2\in\C$ are determined by the initial condition. We note coupling between the $\pm k$ Fourier modes. We also see that a component of the $k$ Fourier mode grows exponentially as $e^{|k|t/2}$, implying that the linear evolution problem is highly unstable, and ill-posed \cite{BM}. Indeed, the instability grows in wavenumber, a fact that some authors describe as a common theme of instability of vortex sheets \cite{protas2021finite}. The authors of \cite{BM} go on to show that there is a family of initially analytic solutions to Equation \eqref{e:hilbert_form_lin} which develop a singularity in finite time. 

\section{Circular vortex sheets}\label{s:circular_sheet}

Now we consider the circular vortex sheet corresponding to the flow field in Equation \ref{e:circ_flow_field}. This sheet is described by the curve 
\begin{equation}\label{e:circularsheet_defn}
    z(\alpha,t)=R(t)e^{i\alpha}(1+\xi(\alpha,t)),~z:[0,2\pi]\times [t_0,t_1] \to  \C \sim \R^2.
\end{equation}
Here $\xi(\alpha,t)$ is a smooth, $2\pi$-periodic perturbation with $||\xi||_{C^2}\ll1$, and $R:[t_0,t_1]\to  \C \sim \R^2$ is a uniform scaling and rotation of the vortex sheet. This ansatz is equivalent to the form used by \cite{Moore1974,SaffmanVortex}; the fact that the perturbation multiplies the vortex sheet merely shifts $\xi$ by one Fourier frequency, a mathematical convenience to decouple the Fourier modes.

\subsection{Steady solution}

Seeking a steady solution to the Birkhoff-Rott equation on which to base our stability analysis, we first set $\xi\equiv0$. Then our starting point from the Birkoff-Rott Equation \eqref{e:bre} is
\begin{align}
    \partial_tz^*(\alpha,t)=\partial_tR^*(t)e^{-i\alpha}&=\frac{1}{2\pi i}\pv\int_0^{2\pi} \frac{1}{R(t)e^{i\alpha}-R(t)e^{i\beta}}d\beta\\
    &=\frac{e^{-i\alpha}}{2\pi iR(t)}\pv\int_0^{2\pi} \frac{1}{1-e^{i(\beta-\alpha)}}d\beta.
\end{align}

In order to simplify our expressions, we will utilize the following generalized formula for the principal value of the Cauchy integral \cite{Blaya2015CauchyIA,grafakos}, which also holds for points on the boundary of the domain after utilizing principle values: we state it here concretely for convenience of the reader.

\begin{lemma}[Cauchy integral formula for Jordan curves]\label{lem:Cauchy-PV}
Let $\Gamma$ be a rectifiable, closed, positively oriented Jordan curve in $\C$ which defines the interior region $\Omega^-$ and exterior region $\Omega^+$. Let $f:\Omega^-\to\C$ be an analytic function and $w_0\in\C/\Gamma$. The well-known Cauchy integral formula states
\begin{equation}
\frac{1}{2\pi i}\pv\int_\Gamma\frac{f(w)}{w-w_0}dw = \begin{cases} f(w_0) &\text{ for } w_0\in\Omega^-\\
 0 &\text{ for } w_0\in\Omega^+
\end{cases}.
\end{equation}
Now suppose $w_0\in\Gamma$; then the Cauchy integral of $f$ at $w_0$ takes the principal value
\begin{equation}\frac{1}{2\pi i}\pv\int_\Gamma\frac{f(w)}{w-w_0}dw = \frac{f(w_0)}{2}.
\end{equation}
\end{lemma}

Continuing our computation from above, we change variables with $w=e^{i(\beta-\alpha)}$, denoting the positively-oriented (counter-clockwise) unit circle by $\partial B$. Then we perform a partial fraction expansion before applying the Cauchy integral formula and its principal value as given in Lemma \ref{lem:Cauchy-PV}:
\begin{align}
    &=\frac{-e^{-i\alpha}}{2\pi R(t)}\pv\int_{\partial B} \frac{1}{w(1-w)}dw\\
    &=\frac{-e^{-i\alpha}}{2\pi R(t)}\pv\int_{\partial B} \frac{1}{w}-\frac{1}{w-1}dw\\
    &=\frac{-e^{-i\alpha}}{2\pi R(t)}(2\pi i -\pi i)\\
    &=-i/2e^{-i\alpha}\frac{1}{R(t)}.
\end{align}
Equating with the left-hand side, we see that the Birkoff-Rott equation is solved by $z(\alpha,t)=R(t)e^{i\alpha}$ when the radius satisfies
\begin{equation}
    R(t)\partial_tR^*(t)=-i/2.\label{e:radius_eq}
\end{equation}
It can be easily seen that $R(t)=e^{it/2}$ satisfies this relation, so a steadily rotating circular vortex sheet is an equilibrium solution. This steady rotation contrasts the steady flat sheet $z(\alpha,t)\equiv\alpha$, which has no time dependence. A more general solution is
\begin{equation}\label{e:radius_form}
    R(t)=R_0e^{it/(2R_0^2)},
\end{equation}
indicating that circular vortex sheets of larger radius rotate more slowly than smaller circular sheets with the same vorticity density. We also note that the form of the expanding sheet used in \cite{Moore1974,SaffmanVortex}:
\begin{equation}\label{e:mooreRadiusform}
    R(t)=R_0(at+1)^n
\end{equation}
cannot satisfy Equation \eqref{e:radius_eq}.

\subsection{Linear stability analysis}

Now we begin a similar computation, applying a small perturbation by allowing $\xi$ to be nonzero. Our starting point is again the Birkoff-Rott Equation \eqref{e:bre}, with the form of circular vortex sheet given by Equation \eqref{e:circularsheet_defn}. First we evaluate the left and right hand sides:
\begin{align}
    \partial_tz^*(\alpha,t)&=\partial_tR^*(t)e^{-i\alpha}(1+\xi^*(\alpha,t))+R^*(t)e^{-i\alpha}\partial_t\xi^*(\alpha,t)\label{e:LHS_full}\\
    &=\frac{1}{2\pi i}\pv\int_0^{2\pi} \frac{1}{R(t)e^{i\alpha}(1+\xi(\alpha,t))-R(t)e^{i\beta}(1+\xi(\beta,t))}d\beta\\
    &=\frac{1}{2\pi iR(t)}\pv\int_0^{2\pi} \frac{1}{e^{i\alpha}-e^{i\beta}}\frac{1}{1+\mathcal{E}}d\beta,
\end{align}
where the quantity
\begin{equation}
\mathcal{E}=\frac{e^{i\alpha}\xi(\alpha,t)-e^{i\beta}\xi(\beta,t)}{e^{i\alpha}-e^{i\beta}}\ll1
\end{equation}
by the smallness of $\xi$. We then Taylor expand to first order about $\mathcal{E}=0$ as
\begin{equation}\label{e:taylorexp1}
\frac{1}{2\pi iR(t)}\pv\int_0^{2\pi} \frac{1}{e^{i\alpha}-e^{i\beta}}-\frac{e^{i\alpha}\xi(\alpha,t)-e^{i\beta}\xi(\beta,t)}{(e^{i\alpha}-e^{i\beta})^2}d\beta.
\end{equation}
Using the left hand side \eqref{e:LHS_full} to eliminate the steady solution found in the previous subsection, we obtain the equation
\begin{multline}
\partial_tR^*(t)e^{-i\alpha}\xi^*(\alpha,t)+R^*(t)e^{-i\alpha}\partial_t\xi^*(\alpha,t)\\
=-\frac{1}{2\pi iR(t)}\pv\int_0^{2\pi}\frac{e^{i\alpha}\xi(\alpha,t)-e^{i\beta}\xi(\beta,t)}{(e^{i\alpha}-e^{i\beta})^2}d\beta.
\end{multline}
Note that this, and any subsequent equalities, neglect terms that are second-order or higher in $\xi$. Using Equation \eqref{e:radius_eq}, we rearrange as
\begin{equation}\label{e:taylorexp_rearr}
    -\pi e^{-i\alpha}\xi^*(\alpha,t)-2\pi i R(t)R^*(t)e^{-i\alpha}\partial_t\xi^*(\alpha,t)
    =\pv\int_0^{2\pi}\frac{e^{i\alpha}\xi(\alpha,t)-e^{i\beta}\xi(\beta,t)}{(e^{i\alpha}-e^{i\beta})^2}d\beta.
\end{equation}
At this point we must consider the form of $\xi$ to proceed. Anticipating coupling between $\pm k$ Fourier modes as well as the need to integrate positive and negative modes separately, we write $\xi(\alpha,t)=A_k(t)e^{ik\alpha}+B_k(t)e^{-ik\alpha}$ where $k>0$. Building on our initial assumptions for $\xi$, we require $||\xi(\alpha,0)||_{C^2}\ll1/(k+1)$ to avoid a singularity of curvature in the initial data. Then the right hand side of Equation \eqref{e:taylorexp_rearr} becomes
\begin{equation}
    A_k(t)\pv\int_0^{2\pi}\frac{e^{i(k+1)\alpha}-e^{i(k+1)\beta}}{(e^{i\alpha}-e^{i\beta})^2}d\beta + B_k(t)\pv\int_0^{2\pi}\frac{e^{-i(k-1)\alpha}-e^{-i(k-1)\beta}}{(e^{i\alpha}-e^{i\beta})^2}d\beta
\end{equation}
\begin{multline}
    = A_k(t)e^{i(k-1)\alpha}\pv\int_0^{2\pi}\frac{1-e^{i(k+1)(\beta-\alpha)}}{(1-e^{i(\beta-\alpha)})^2}d\beta \\
    + B_k(t)e^{-i(k+1)\alpha}\pv\int_0^{2\pi}\frac{1-e^{-i(k-1)(\beta-\alpha)}}{(1-e^{i(\beta-\alpha)})^2}d\beta.
\end{multline}
For the first term (positive Fourier modes), we change variables with $w_1=e^{i(\beta-\alpha)}$ to the positively oriented unit circle $\partial B$. For the second term with negative Fourier modes, we apply the change of variables $w_2=e^{-i(\beta-\alpha)}$ to the negatively oriented unit circle $\partial B^-$, and then flip direction of integration to the positively oriented unit circle $\partial B$; this contributes a negative sign.\footnote{In \cite{RyanNonlinear}, we failed to separately handle the negative Fourier modes, resulting in a critical sign error; here we correct the mistake by changing variables separately. The details are discussed in \cite{Ryancorrigendum}.} The transformed right hand side is
\begin{multline}
    -iA_k(t)e^{i(k-1)\alpha}\pv\int_{\partial B} \frac{1-w_1^{k+1}}{w_1(1-w_1)^2}dw_1 \\
    -iB_k(t)e^{-i(k+1)\alpha}\pv\int_{\partial B} \frac{1-w_2^{k-1}}{w_2(1-1/w_2)^2}dw_2.
\end{multline}
We apply a telescoping sum and partial fraction expansion to set up for integration.
\begin{multline}
    -iA_k(t)e^{i(k-1)\alpha}\sum_{j=0}^{k}\pv\int_{\partial B} \frac{w_1^j}{w_1(1-w_1)}dw_1 \\
    -iB_k(t)e^{-i(k+1)\alpha}\sum_{j=0}^{k-2}\pv\int_{\partial B} \frac{w_2^j(1-w_2)}{w_2(1-1/w_2)^2}dw_2
\end{multline}
\begin{multline}
    =-iA_k(t)e^{i(k-1)\alpha}\sum_{j=0}^{k}\pv\int_{\partial B} \frac{w_1^j}{w_1} - \frac{w_1^j}{w_1-1}dw_1 \\
    + iB_k(t)e^{-i(k+1)\alpha}\sum_{j=0}^{k-2}\pv\int_{\partial B} w_2^j + \frac{w_2^j}{w_2-1} dw_2.
\end{multline}
Now we can integrate using the Cauchy integral theorem and Lemma \ref{lem:Cauchy-PV}. Note that the first term of the first integral is everywhere analytic only for $j>0$. Performing the integration and substituting, the above expression is equal to
\begin{align}
    &-iA_k(t)e^{i(k-1)\alpha}\left(2\pi i - \sum_{j=0}^{k}\pi i\right) +iB_k(t)e^{-i(k+1)\alpha}\sum_{j=0}^{k-2} \pi i\\
    &=iA_k(t)e^{i(k-1)\alpha}(k-1)\pi i +iB_k(t)e^{-i(k+1)\alpha}(k-1)\pi i\\
    &=-A_k(t)\pi(k-1)e^{i(k-1)\alpha}-B_k(t)\pi(k-1)e^{-i(k+1)\alpha}.
\end{align}
The left hand side is $-\pi e^{-i\alpha}\xi^*(\alpha,t)-2\pi i R(t)R^*(t)e^{-i\alpha}\partial_t\xi^*(\alpha,t)$. We expand $\xi$, equate with the right hand side, and eliminate $e^{-i\alpha}$ to obtain
\begin{multline}\label{e:full_lin_ev}
    A_k^*(t)e^{-ik\alpha}+B_k^*(t)e^{ik\alpha} + 2iR(t)R^*(t)\left(\partial_t A_k^*(t)e^{-ik\alpha}+\partial_t B_k^*(t)e^{ik\alpha}\right)\\
    = A_k(t)(k-1)e^{ik\alpha}+B_k(t)(k-1)e^{-ik\alpha}.
\end{multline} 
Equation \eqref{e:full_lin_ev} is a linear differential equation for the Fourier coefficients $A_k$ and $B_k$. We note coupling between $\pm$k Fourier modes, as in the flat case. If we had instead used the form $z(\alpha,t)=R(t)(e^{i\alpha}+\xi(\alpha,t))$, where the perturbation is simply added to the steady state, coupling between all Fourier frequencies would emerge, and the system could not be closed without neglecting low-frequency terms. This full coupling of the frequency domain could result in an instability which manifests very rapidly at all spatial scales, suggesting an immediate transition to turbulence. However, the form we used here is preferred, because it allows us to precisely examine the stability of the full linearized equation without neglecting any terms.

To examine the stability of Equation \eqref{e:full_lin_ev} we must substitute for $R(t)$ the steady form given by Equation \eqref{e:radius_form}. Then the evolution equation becomes
\begin{multline}\label{e:full_lin_ev_radius}
    R_0^2\left(\partial_t A_k^*(t)e^{-ik\alpha}+\partial_t B_k^*(t)e^{ik\alpha}\right) = \\ 
    -i/2\left(A_k(t)(k-1)e^{ik\alpha}+B_k(t)(k-1)e^{-ik\alpha} - A_k^*(t)e^{-ik\alpha} - B_k^*(t)e^{ik\alpha}\right).
\end{multline} 
Matching Fourier coefficients we obtain
\begin{align}
    &R_0^2\partial_t A_k^*(t) =  
    i/2\left(A_k^*(t)-B_k(t)(k-1)\right),\\
    &R_0^2\partial_t B_k^*(t) =  
    i/2\left(B_k^*(t)-A_k(t)(k-1)\right).
\end{align}

\pagebreak

Breaking the Fourier modes up into real and imaginary parts as $A_k(t)=A_k^R(t)+iA_k^I(t)$, $B_k(t)=B_k^R(t)+iB_k^I(t)$, the system above can be written in matrix form:
\begin{equation}\label{e:matrix_system}
{
\renewcommand{\arraystretch}{1.5}  
\partial_t \begin{bmatrix} A^R_k(t)\\A^I_k(t)\\B_k^R(t)\\B_k^I(t) \end{bmatrix} = \begin{bmatrix} 0 & \frac{1}{2R_0^2} & 0 & \frac{k-1}{2R_0^2} \\ \frac{-1}{2R_0^2} & 0 & \frac{k-1}{2R_0^2} & 0 \\ 0 & \frac{k-1}{2R_0^2} & 0 & \frac{1}{2R_0^2} \\ \frac{k-1}{2R_0^2} & 0 & \frac{-1}{2R_0^2} & 0 \end{bmatrix}\begin{bmatrix} A^R_k(t)\\A^I_k(t)\\B_k^R(t)\\B_k^I(t) \end{bmatrix}.
}
\end{equation}
This is a homogeneous system of linear differential equations eigenvalues at $\lambda_{1,2}= -\sqrt{(k-2)k}/(2R_0^2)$ and $\lambda_{3,4}= \sqrt{(k-2)k}/(2R_0^2) $, implying that for $k>2$ the Fourier coefficients grow exponentially, i.e., the linear evolution problem is unstable. The only stable modes are $k=0$ (corresponding to dilations and time rescaling, and represented by $R$ rather than $\xi$), $k=1$ (corresponding to deformation and translation) and $k=2$ (corresponding to an oval-shaped deformation, not elliptical). The stability of these low modes is a consequence of the circular geometry, which necessitates a time-dependent radius, contributing an additional term to the left hand side which is not present in the flat case. The form of exponential growth can be clearly seen by diagonalizing the matrix in \eqref{e:matrix_system} about its eigenvalues to construct the general solution
\begin{equation}\label{e:matrix_soln}
{
\renewcommand{\arraystretch}{1.5}  
\begin{bmatrix} A^R_k(t)\\A^I_k(t)\\B_k^R(t)\\B_k^I(t) \end{bmatrix} = \begin{bmatrix} \frac{\sqrt{(k-2)k}}{1-k} & \frac{1}{k-1} & \frac{\sqrt{(k-2)k}}{k-1} & \frac{1}{k-1} \\ \frac{1}{1-k} & \frac{\sqrt{(k-2)k}}{1-k} & \frac{1}{1-k} & \frac{\sqrt{(k-2)k}}{k-1} \\ 0 & 1 & 0 & 1 \\ 1 & 0 & 1 & 0 \end{bmatrix}\begin{bmatrix} c_1e^{\lambda_1t}\\c_2e^{\lambda_2t}\\c_3e^{\lambda_3t}\\c_4e^{\lambda_4t} \end{bmatrix},
}
\end{equation}
where $c_i$ are constants determined by the initial condition. The important thing to note here is that the $k$th Fourier mode has a component that grows like $e^{\sqrt{(k-2)k}/(2R_0^2)t}$. As is characteristic of the Kelvin-Helmholtz instability, the strength of instability is enhanced at short wavelengths; however, the strength of instability is weaker at any given wavenumber than for the flat sheet. Another distinguishing feature of the circular case is that, for the strongest instability, wavenumber must be large relative to the radius of the vortex sheet; conversely, smaller circular sheets with the same strength become unstable more rapidly. Equivalently by rescaling space and time, stronger vortex sheets of a given radius have a stronger instability. 

Although a steady expanding solution has not yet been identified, this inverse relationship between growth rate and radius suggests that expansion of the circular sheet, as if by a source at the origin, may suppress the Kelvin-Helmholtz instability. This notion is in agreement with theory for the flat sheet, in which case a local increase in length faster than $t^{1/2}$ stabilizes the sheet \cite{saffman_1979,SaffmanVortex}.

To summarize the instability of circular vortex sheets, we state our main theorem. 

\begin{theorem}
    \label{thm:main_instab}
    Consider the perturbed circular vortex sheet $z(\alpha,t)=R(t)e^{i\alpha}(1+\xi(\alpha,t))$, $z:[0,2\pi]\times [t_0,t_1] \to  \C \sim \R^2$, with $R(t)=R_0e^{it/(2R_0^2)}$ and $\xi(\alpha,t)=A_k(t)e^{ik\alpha}+B_k(t)e^{-ik\alpha}$ where initially $||\xi||_{C^2}\ll1$. When linearized about the steady solution $\xi\equiv0$, the $k$th Fourier mode of $\xi$ has a component that grows like $e^{\sqrt{(k-2)k}/(2R_0^2)t}$; thus for $k>2$ the circular sheet exhibits the Kelvin-Helmholtz instability.
\end{theorem}

It is insightful to compare our result with the growth rate obtained by \cite{Moore1974,SaffmanVortex} for the radius form given by Equation \eqref{e:mooreRadiusform}. By introducing the dimensionless time $\tau=at+1$ and defining
\begin{equation}
    \beta=\Gamma\sqrt{(k-2)k}/(2\pi aR_o^2),
\end{equation}
where $\Gamma$ is a constant related to the strength of the vortex sheet $\Gamma/(2\pi R)$, the authors obtain the general solution 
\begin{equation}\label{e:gensolnMoore}
    \eta(t)=\tau^{1/2}\left(CI_{|p|}(\beta|p|\tau^{1/(2p)}) + DK_{|p|}(\beta|p|\tau^{1/(2p)})\right).
\end{equation}
Here $I_\alpha$ and $K_\alpha$ are modified Bessel functions of the first and second kind, respectively, $C$ and $D$ are constants of integration, and $\eta$ is a transformation (related to the radius of the vortex sheet) of the perturbation quantity $\eps$. The modified Bessel functions may be simplified in the case of half-integer order \cite{Mainardi2022}. For the non-expanding vortex sheet, corresponding to $n=0$ and $p=1/2$ in the notation of \cite{Moore1974,SaffmanVortex}, Equation \eqref{e:gensolnMoore} reduces to
\begin{equation}\label{e:MooreReduced_soln}
    \eta(t)=C(\pi\beta)^{-1/2}e^{\beta\tau/2}+(D(\pi/\beta)^{1/2}-C(\pi\beta)^{-1/2})e^{-\beta\tau/2}.
\end{equation}
Therefore there is a component of the perturbation which grows like
\begin{equation}
    |\eps(t)| \sim e^{\Gamma\sqrt{(k-2)k}/(4\pi aR_o^2)\tau}.
\end{equation}
Thus the spectrum found by \cite{Moore1974,SaffmanVortex} matches what we present in Theorem \ref{thm:main_instab}, up to rescaling by the vortex sheet strength. Our method of proof, rather than the resulting exponential growth rate, is the novel contribution of this paper. By providing a fully explicit solution in Equation \eqref{e:matrix_soln}, we lay a strong foundation for further results. In the next section, we use this foundation to examine linear singularity formation.

\subsection{Singularity formation for the linearized equation}


With explicit solutions to the linearized equation given by Equation \eqref{e:matrix_soln}, we can construct initially analytic data which form a singularity in finite time. Consider the perturbation initially defined by
\begin{equation}
    \xi(\alpha,0)=\varepsilon\sum_{k>2}(A_k^R(0)+iA_k^I(0))e^{ik\alpha}+(B_k^R(0)+iB_k^I(0))e^{-ik\alpha},
\end{equation}
where $\varepsilon \ll 1$. We consider the form
\begin{equation}
    A_k^R(0)=e^{-k}e^{-1/\Gamma\sqrt{(k-2)k}},
\end{equation}
where $\Gamma>0$ is a constant representing the vortex sheet strength and $A_k^I(0)=B_k^R(0)=B_k^I(0)=0$. From Equation \eqref{e:matrix_soln}, we have the constants
\begin{align}
    &c_1=-e^{-k}e^{-1/\Gamma\sqrt{(k-2)k}}\left(\frac{(k-1)\sqrt{(k-2)k}}{2(k-2)k}\right),\\
    &c_3=e^{-k}e^{-1/\Gamma\sqrt{(k-2)k}}\left(\frac{(k-1)\sqrt{(k-2)k}}{2(k-2)k}\right),\\
    &c_2=c_4=0.
\end{align}
This form guarantees that the Fourier series converges to analytic data for at least some time \cite{BM, moore_singularity, strang_cse}. Then from Equation \eqref{e:matrix_soln} the positive real mode evolves like
\begin{equation}
    A_k^R(t)=1/2e^{-k}e^{-1/\Gamma\sqrt{(k-2)k}}e^{-\sqrt{(k-2)k}/(2R_0^2)t}+1/2e^{-k}e^{-1/\Gamma\sqrt{(k-2)k}}e^{\sqrt{(k-2)k}/(2R_0^2)t}.
\end{equation}
The first term in this expression decays with time, but the second grows without bound. The growth rate of the second term gives rise to the inequality
\begin{equation}
    -k+(t/(2R_0^2)-1/\Gamma)\sqrt{(k-2)k}\leq-k
\end{equation}
which, when satisfied, says that Fourier modes decay at least exponentially in $k$ and so the solution remains analytic. This inequality fails after the time
\begin{equation}
    t=\tau=\frac{2R_0^2}{\Gamma}.
\end{equation}
At this point the Fourier modes decay less than exponentially in $k$; thus the Fourier series no longer converges to an analytic function, and a singularity has formed. Due to this singularity, the solution after $t=\tau$ no longer depends continuously on its initial data, implying ill-posedness of the linear evolution problem as discussed in \cite{BM}. As we anticipated, the singularity develops more rapidly for sheets of smaller radius or, equivalently, greater strength. We summarize this result with the following proposition. 

\begin{proposition}
    \label{prop:approx_sing}
    Define the perturbed circular vortex sheet as in Theorem \ref{thm:main_instab}; then for $k>2$ the linear equation \eqref{e:full_lin_ev_radius} develops a singularity from analytic initial data in a time proportional to the square of the radius.
\end{proposition}

\section{Nonlinear simulations}\label{s:numerical_demos}

In this section, we will demonstrate that the linear solution given by Equation \eqref{e:matrix_soln} is sufficient to describe the wave-breaking mechanism of the Kelvin-Helmholtz instability. For a point of comparison, we simulate the full nonlinear evolution of the Birkhoff-Rott equation. Well-established numerical methods exist to simulate vortex sheet evolution up to and beyond the development of singularity \cite{krasny1986study,krasny1990computing,sun2012generalized,Sohn2013}, particularly including the point vortex method. Given a vortex sheet parameterized by vorticity, we consider the discretization $z_j=z(\alpha_j,t),~j=1,\ldots, N$. By approximating the sheet using point vortices at the points $z_j$, we can derive the system of ordinary differential equations \cite{krasny_2008}
\begin{equation}
    \frac{dz^*_j}{dt}=\frac{1}{2\pi i}\sum_{k=1,k\neq j}^N K(z_j,z_k)w_k,
    \label{e:pvm}
\end{equation}
where the $w_k$ are quadrature weights for the spatial discretization. A natural choice of kernel is
\begin{equation}K(z_1,z_2) = \frac{1}{z_1-z_2},
\end{equation}
which is the kernel associated with the Birkhoff-Rott Equation \eqref{e:bre}, and in the discrete case it views vorticity as concentrated exactly at each $z_j$. Before the time of singularity development \cite{moore_singularity}, the point vortex method with this kernel converges towards the Birkhoff-Rott solution as $N\to\infty$ \cite{krasny1986study}. After roll-up begins, the numerical evolution becomes unstable and kernel must be regularized to proceed further in time. One such regularization is the Krasny kernel \cite{krasny_desing}
\begin{equation}K(z_1,z_2)=\Gamma_{PV}\frac{1}{z_1-z_2}\frac{|z_1-z_2|^2}{|z_1-z_2|^2+\delta^2},
\end{equation}
which can be conveniently rewritten using properties of complex numbers as
\begin{equation}
    K(z_1,z_2) = \Gamma_{PV}\frac{(z_1 - z_2)^*}{|z_1 - z_2|^2 + \delta^2}.
    \label{e:krasnykernel}
\end{equation}
Here $\delta$ is a smoothing parameter which moderates the singularity in the Cauchy kernel; its presence corresponds to vorticity spread over small areas around each $z_j$. The constant $\Gamma_{PV}$ represents the strength of each point vortex. Higher values of $\Gamma_{PV}$ result in more rapid instability. There are a variety of other options for regularized kernels including the $\alpha$-regularization studied by \cite{bardos2010global}. We will use the Krasny kernel here.

As an example to facilitate comparison, we choose a perturbed circular vortex sheet described by $z(\alpha,t)=e^{it/2}e^{i\alpha}(1+\xi(\alpha,t))$. We apply the cosine perturbation $\xi(\alpha,t)=0.02\cos{k\alpha}$ with $k=6$. In all simulations, the small dot initially at $(x_0,y_0)=(1,0)$ tracks the rotation of the vortex sheet.

\begin{figure}[hbpt!]
    \centering
    \begin{subfigure}{.31\textwidth}
        \centering
        \includegraphics[width=\linewidth]{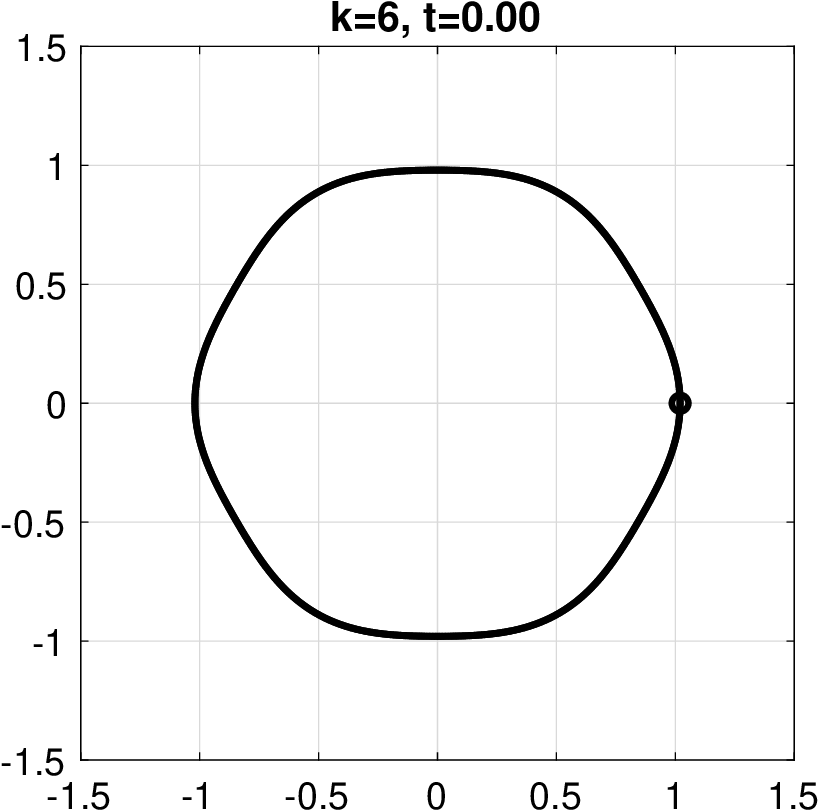}
        \caption{Initial condition.}\label{f:nonlin_sim_a}
    \end{subfigure}
    \begin{subfigure}{.31\textwidth}
        \centering
        \includegraphics[width=\linewidth]{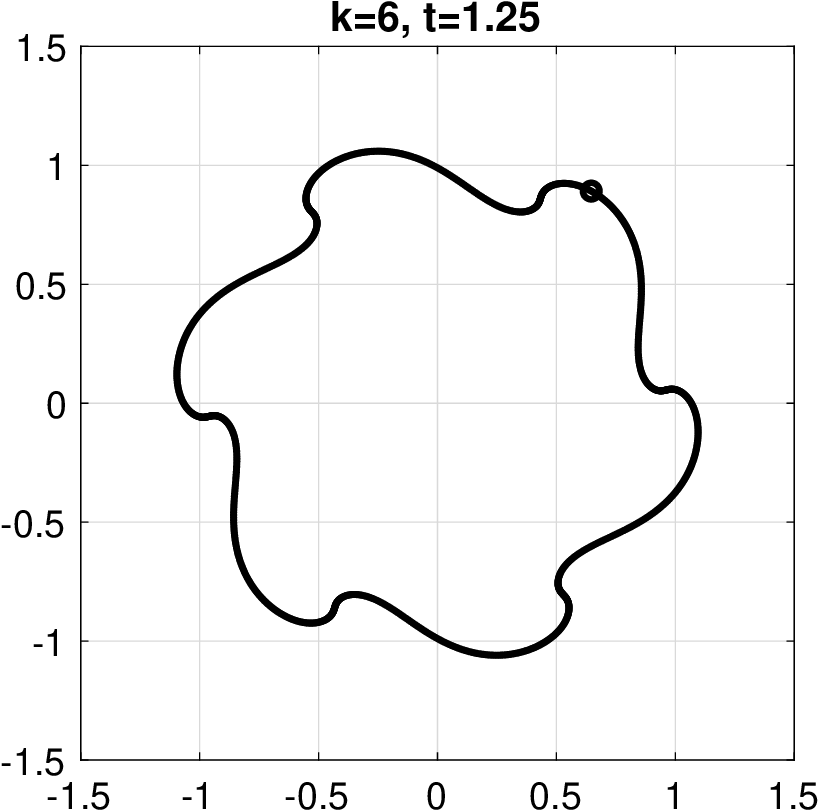}
        \caption{Singularity formation.}\label{f:nonlin_sim_b}
    \end{subfigure}
    \begin{subfigure}{.31\textwidth}
        \centering
        \includegraphics[width=\linewidth]{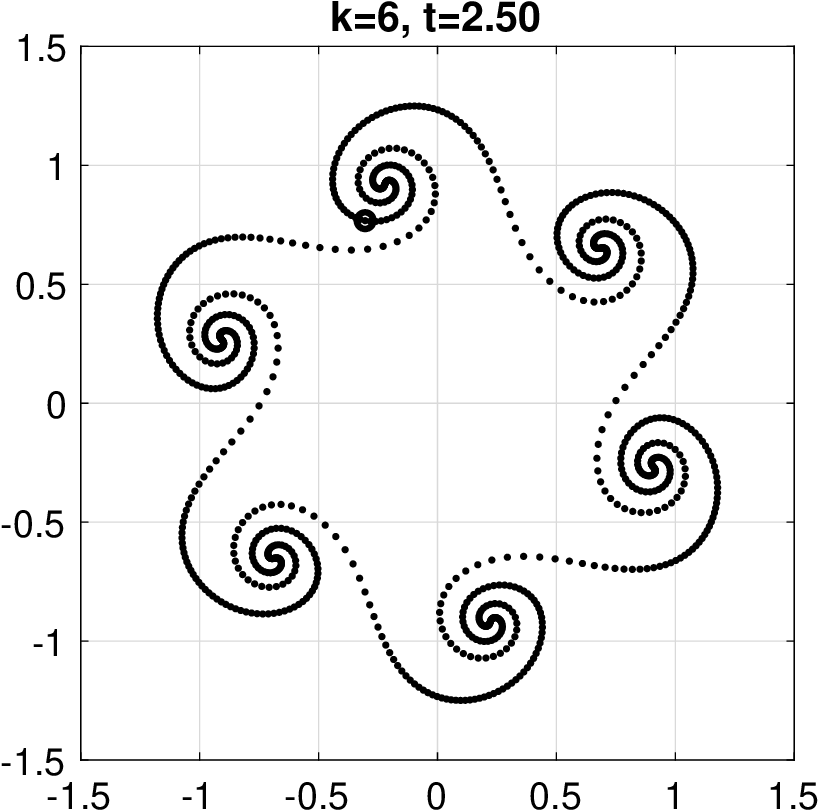}
        \caption{Spiral rollup.}\label{f:nonlin_sim_c}
    \end{subfigure}
    \caption{The full nonlinear Kelvin-Helmholtz instability developing on a circular vortex sheet from an initial cosine perturbation with $k=6$. The second panel \ref{f:nonlin_sim_b} shows the critical moment of instability.}
    \label{f:nonlin_sim}
\end{figure}

Figure \ref{f:nonlin_sim} shows the full nonlinear evolution of this initial condition according to the Birkhoff-Rott equation, calculated by the point vortex method with Krasny regularization, smoothing parameter $\delta=0.1$, and $N=1000$ point vortices. Figure \ref{f:nonlin_sim_b} shows the approximate time where the curve ceases to be a single-valued function of the tangential coordinate; this is regarded as the critical time of wave-breaking instability. Note that we have scaled the strength of point vortices with $\Gamma_{PV}=1.42$ so that the nondimensional time of singularity formation matches the linear solution.

In Figure \ref{f:nonlin_sim_c}, we see well-developed spiral vortices. The development of these spiral vortices, and the extension of the vortex sheet in between the vortices, represents the cascade of squared vorticity (enstrophy) to higher wavenumbers. This mechanism suggests that at further time the vortices could evolve independently, consolidating and developing further instability (though the present numerical scheme does not permit it). In this way, the Kelvin-Helmholtz instability can progressively generate eddies at a range of spatial scales, resulting in a fully turbulent flow.

\begin{figure}[hbpt!]
    \centering
    \begin{subfigure}{.31\textwidth}
        \centering
        \includegraphics[width=\linewidth]{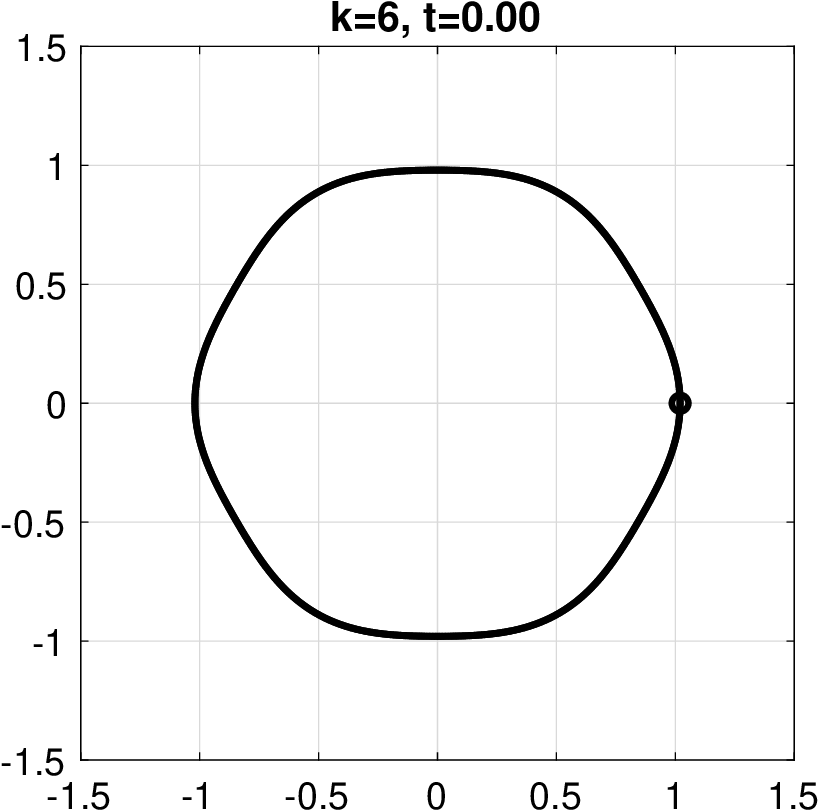}
        \caption{Initial condition.}\label{f:lin_sim_a}
    \end{subfigure}
    \begin{subfigure}{.31\textwidth}
        \centering
        \includegraphics[width=\linewidth]{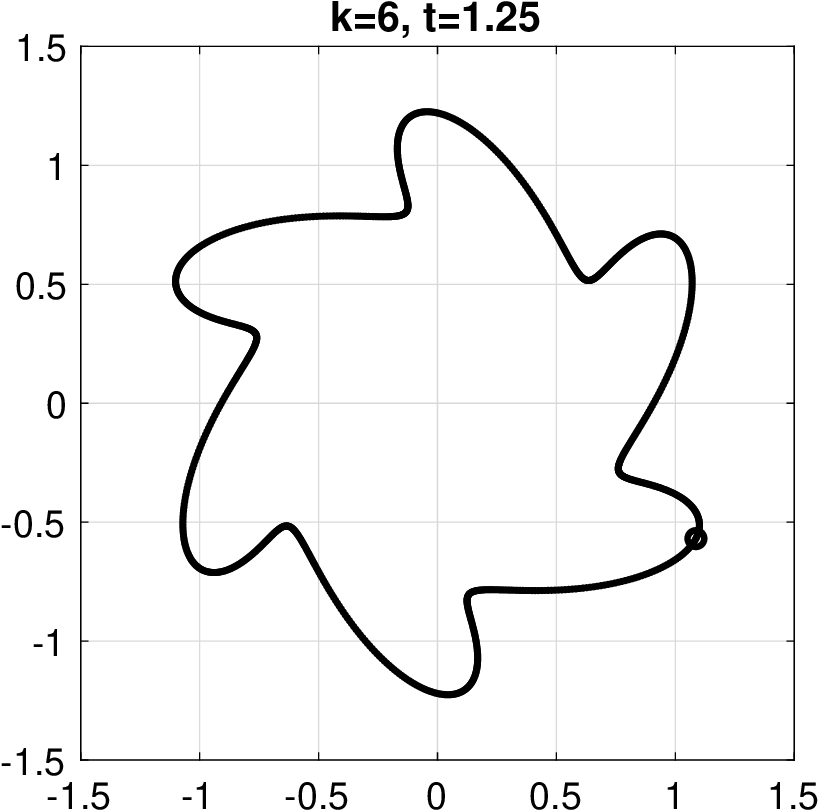}
        \caption{Singularity formation.}\label{f:lin_sim_b}
    \end{subfigure}
    \begin{subfigure}{.31\textwidth}
        \centering
        \includegraphics[width=\linewidth]{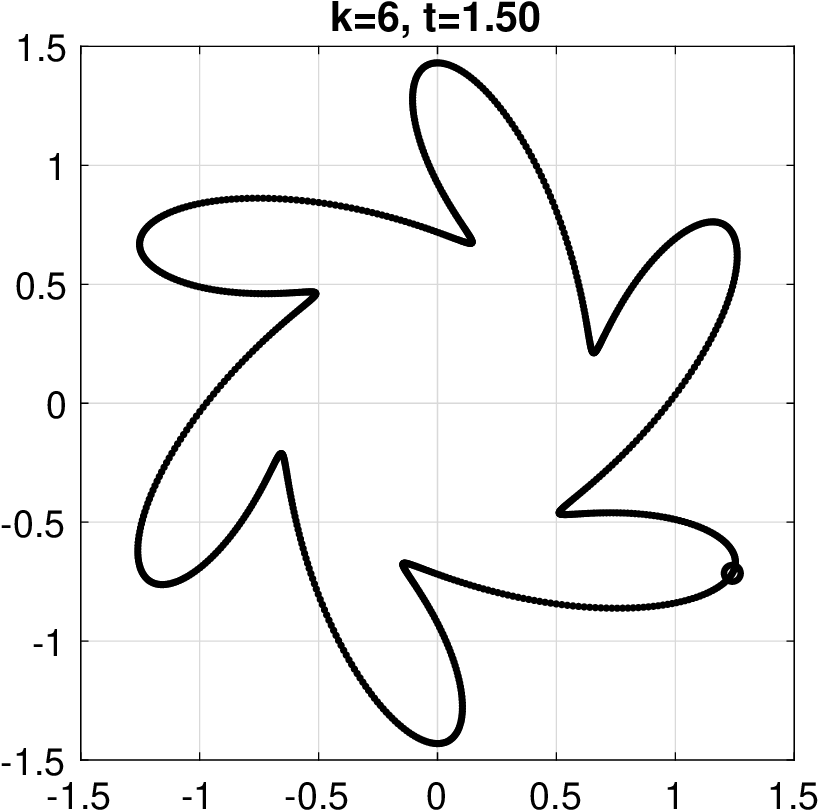}
        \caption{Blowup.}\label{f:lin_sim_c}
    \end{subfigure}
    \caption{The linearized evolution of a circular vortex sheet from an initial cosine perturbation with $k=6$. The second panel \ref{f:lin_sim_b} shows the critical moment of instability. The third panel is included to confirm that the linear dynamics fail to capture spiral roll-up.}
    \label{f:lin_sim}
\end{figure}

If the linear dynamics are sufficient to explain the Kelvin-Helmholtz instability on the circular sheet, we should see the linear evolution reproduce the same mechanism of wave-breaking by developing of a singularity in the tangential derivative of the curve. To see this, we directly implement the algebraic solution given by Equation \eqref{e:matrix_soln}, and reassemble the Fourier components to plot as a single curve. Figure \ref{f:lin_sim} shows the result at three time stills. The initial condition is the same as for the nonlinear simulation.

The shape of the curve is somewhat different from the nonlinear case, but at $t=1.25$ a singularity develops in the tangential derivative, marking the critical moment of wave-breaking instability; this is shown in Figure \ref{f:lin_sim_b}, and suggests that the linear dynamics adequately represent the initial mechanism of Kelvin-Helmholtz instability. Concurrently a sharp corner forms; this is not evident in Figure \ref{f:nonlin_sim}. Continuing the solution is not mathematically justified after this point, but we include Figure \ref{f:lin_sim_c} to show that the linear mechanism fails to reproduce the formation of spiral vortices, confirming that rolling up of the vortex sheet is a nonlinear phenomenon.

\begin{figure}[hbpt!]
    \centering
    \begin{subfigure}{.31\textwidth}
        \centering
        \includegraphics[width=\linewidth]{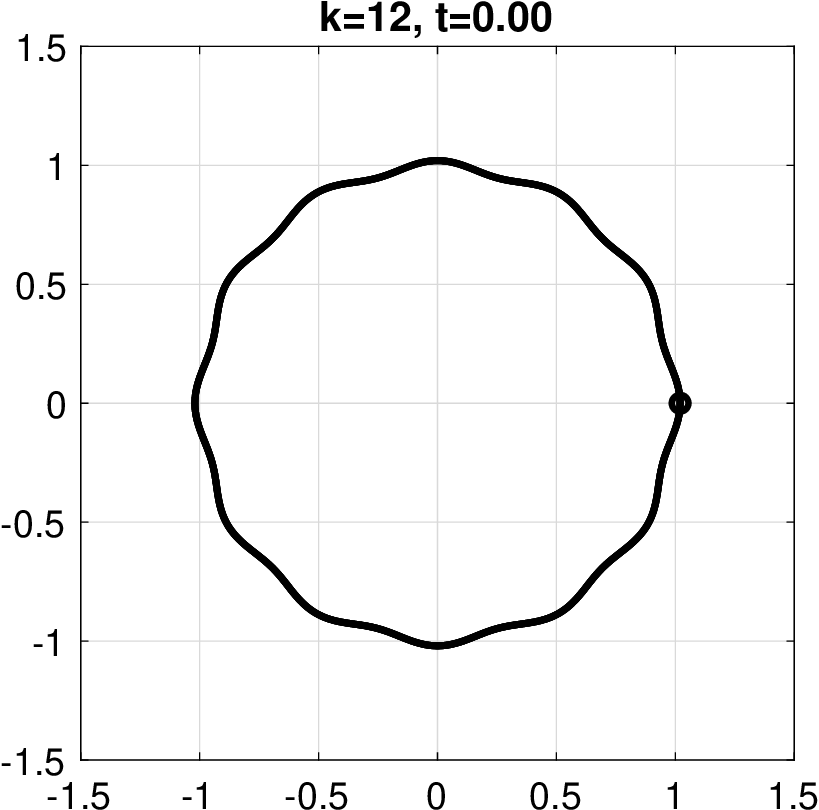}
        \caption{Initial condition.}
    \end{subfigure}
    \begin{subfigure}{.31\textwidth}
        \centering
        \includegraphics[width=\linewidth]{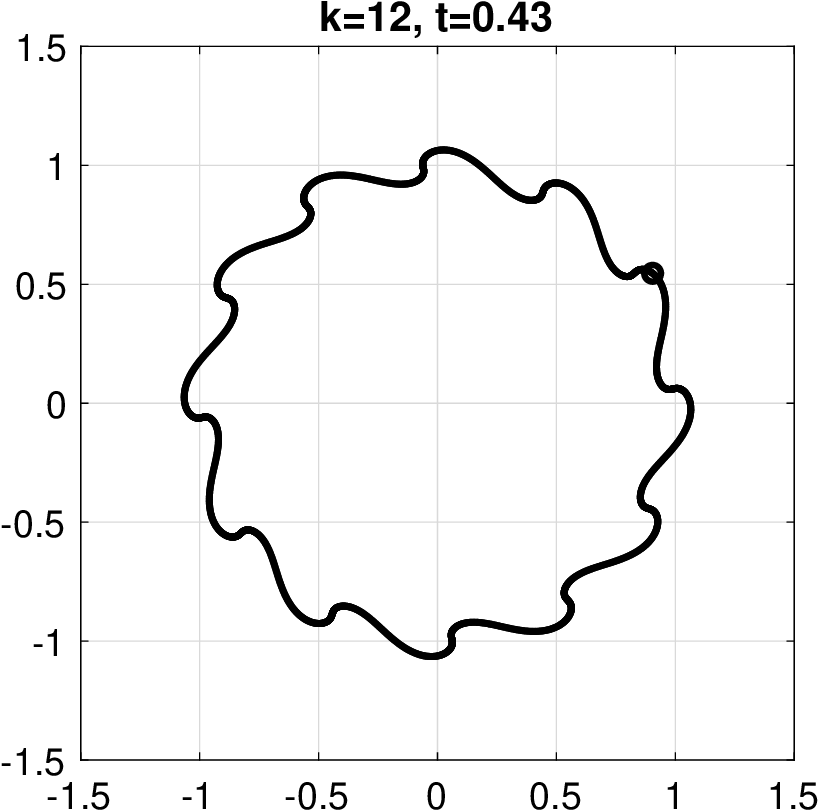}
        \caption{Singularity formation.}\label{f:nonlin12_sim_c}
    \end{subfigure}
    \begin{subfigure}{.31\textwidth}
        \centering
        \includegraphics[width=\linewidth]{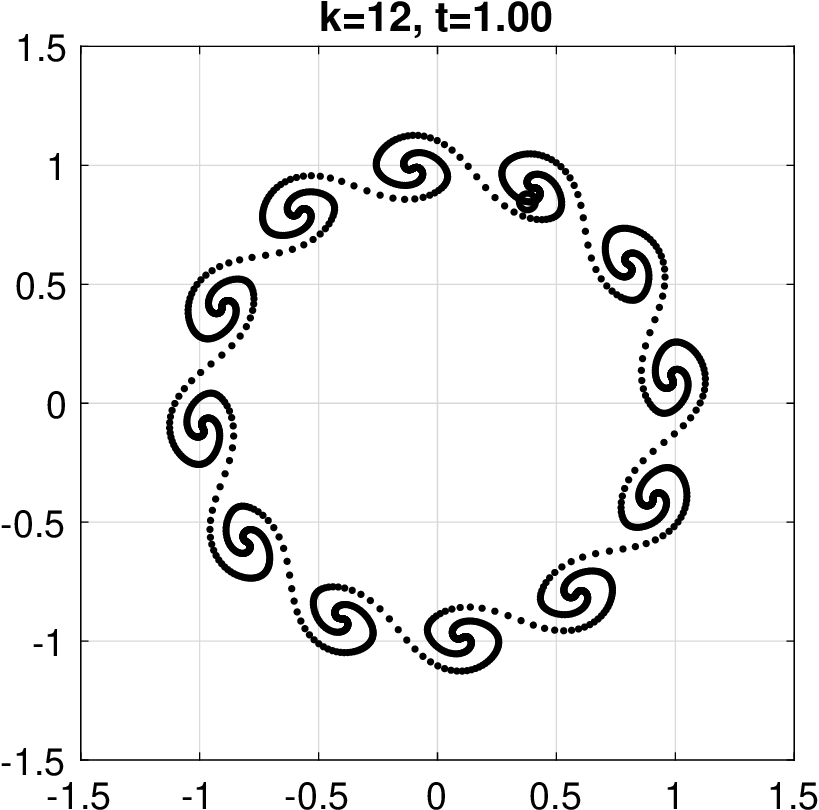}
        \caption{Spiral rollup.}
    \end{subfigure}
    \caption{The full nonlinear Kelvin-Helmholtz instability developing on a circular vortex sheet from an initial cosine perturbation with $k=12$.}
    \label{f:nonlin_sim12}
\end{figure}

\begin{figure}[hbpt!]
    \centering
    \begin{subfigure}{.31\textwidth}
        \centering
        \includegraphics[width=\linewidth]{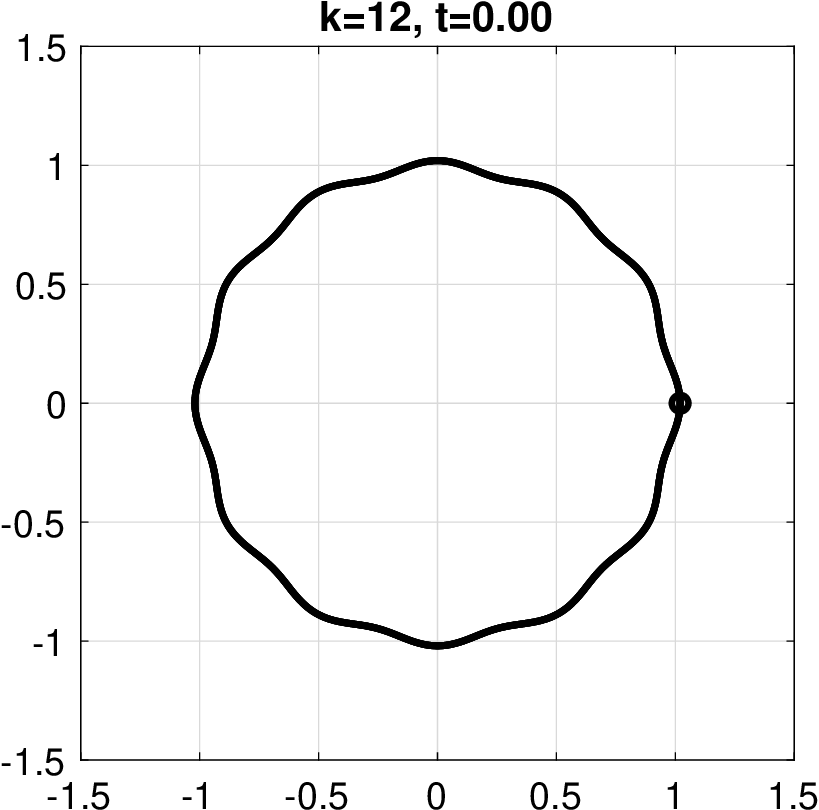}
        \caption{Initial condition.}
    \end{subfigure}
    \begin{subfigure}{.31\textwidth}
        \centering
        \includegraphics[width=\linewidth]{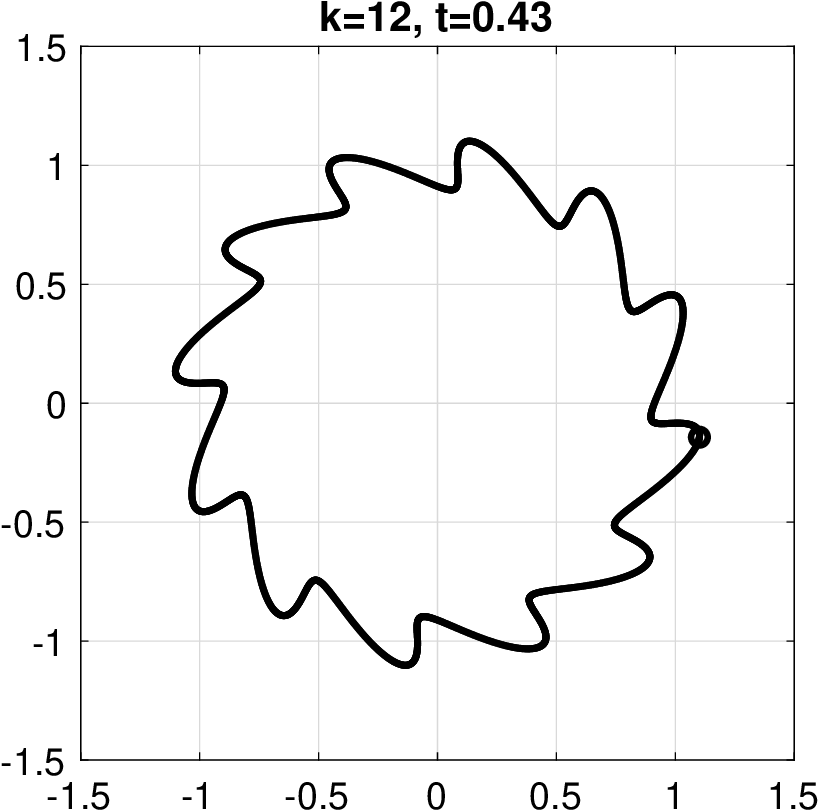}
        \caption{Singularity formation.}
    \end{subfigure}
    \begin{subfigure}{.31\textwidth}
        \centering
        \includegraphics[width=\linewidth]{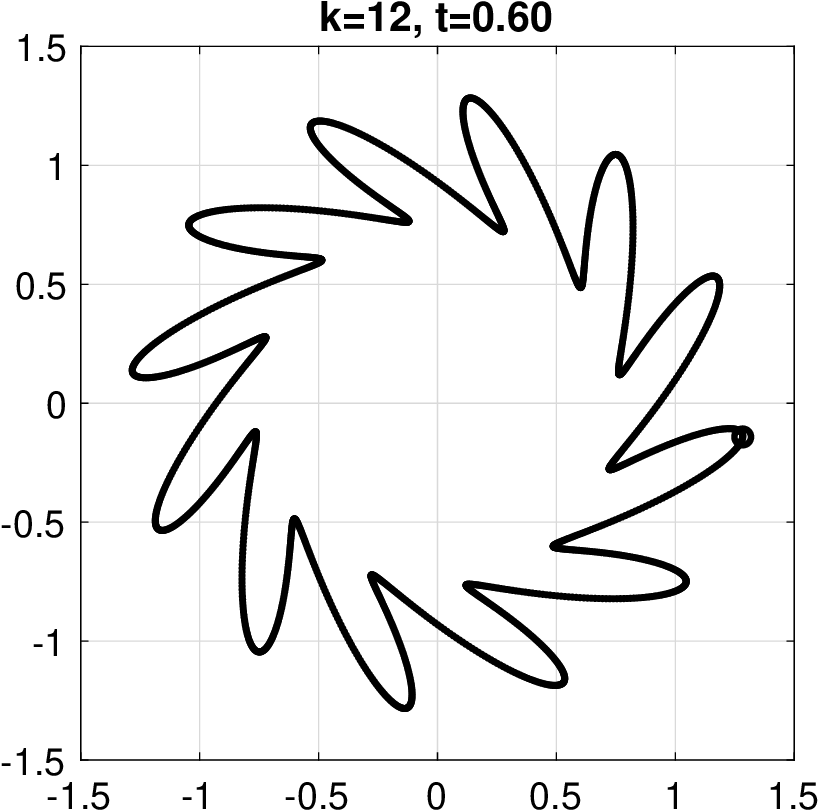}
        \caption{Blowup.}\label{f:lin_sim_c12}
    \end{subfigure}
    \caption{The linearized evolution of a circular vortex sheet from an initial cosine perturbation with $k=12$.}
    \label{f:lin_sim12}
\end{figure}

A higher-frequency perturbation is of greater physical interest, so we repeat our comparison with the cosine perturbation $\xi(\alpha,t)=0.02\cos{k\alpha}$; the new parameters are $k=12$, $N=2000$, and $\Gamma_{PV}=2.25$. As expected, the development of instability is more rapid in both the nonlinear simulation and linear solution. Figure \ref{f:nonlin_sim12} shows the nonlinear evolution, and Figure \ref{f:lin_sim12} shows the linearized evolution. Again, the linear dynamics capture the wave-breaking mechanism of the Kelvin-Helmholtz instability. 

To quantify our comparison of linear and nonlinear dynamics, we plot a time series for the maximum magnitude of the $k$th Fourier mode. We take the maximum perturbation magnitude in the tangential direction as our diagnostic quantity, because there are always points on the vortex sheet that either remain at a fixed radius or travel inwards. In the linear case, we do not have full coupling of the Fourier modes, so the $k$th mode is already known. An inverse Fourier transform is used to extract this mode from the nonlinear simulation. We also calculate the maximum total perturbation magnitude for the nonlinear simulation, to show the effect of nonlinear coupling.

\begin{figure}[hbpt!]
    \centering
    \begin{subfigure}{.465\textwidth}
        \centering
        \includegraphics[width=\linewidth]{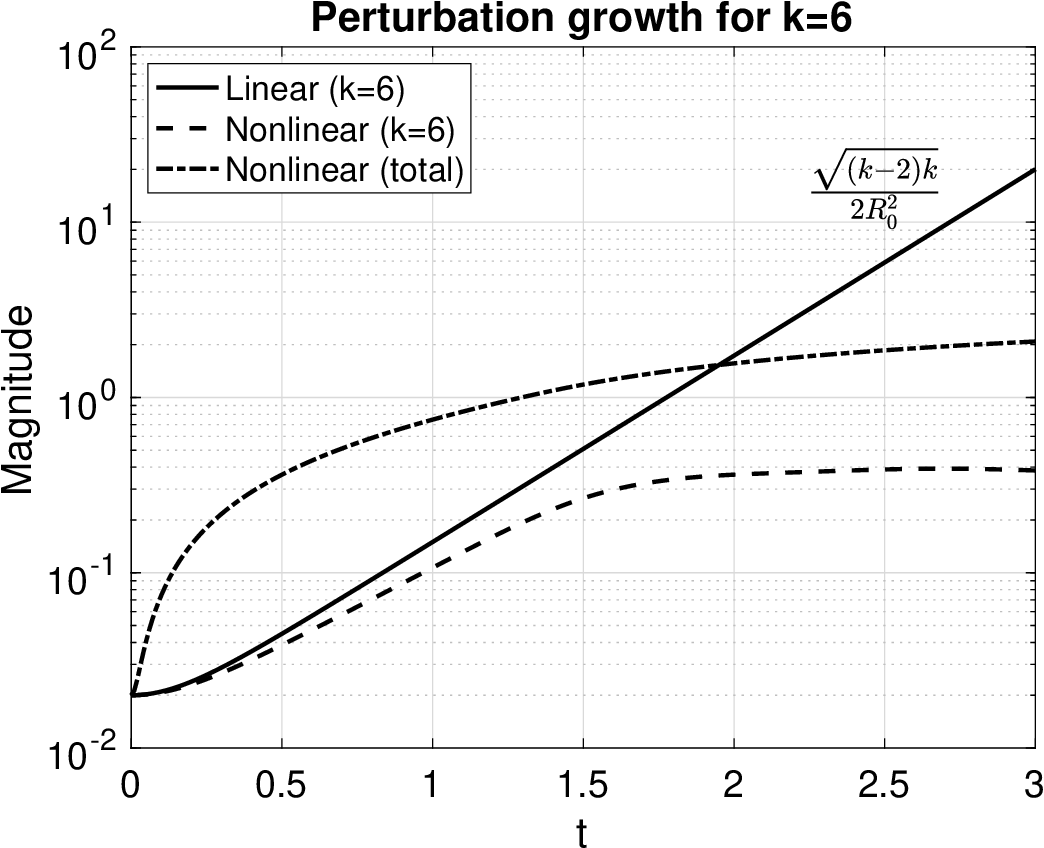}
        \caption{For initial condition with $k=6$.}\label{f:comp_k6}
    \end{subfigure}
    \begin{subfigure}{.465\textwidth}
        \centering
        \includegraphics[width=\linewidth]{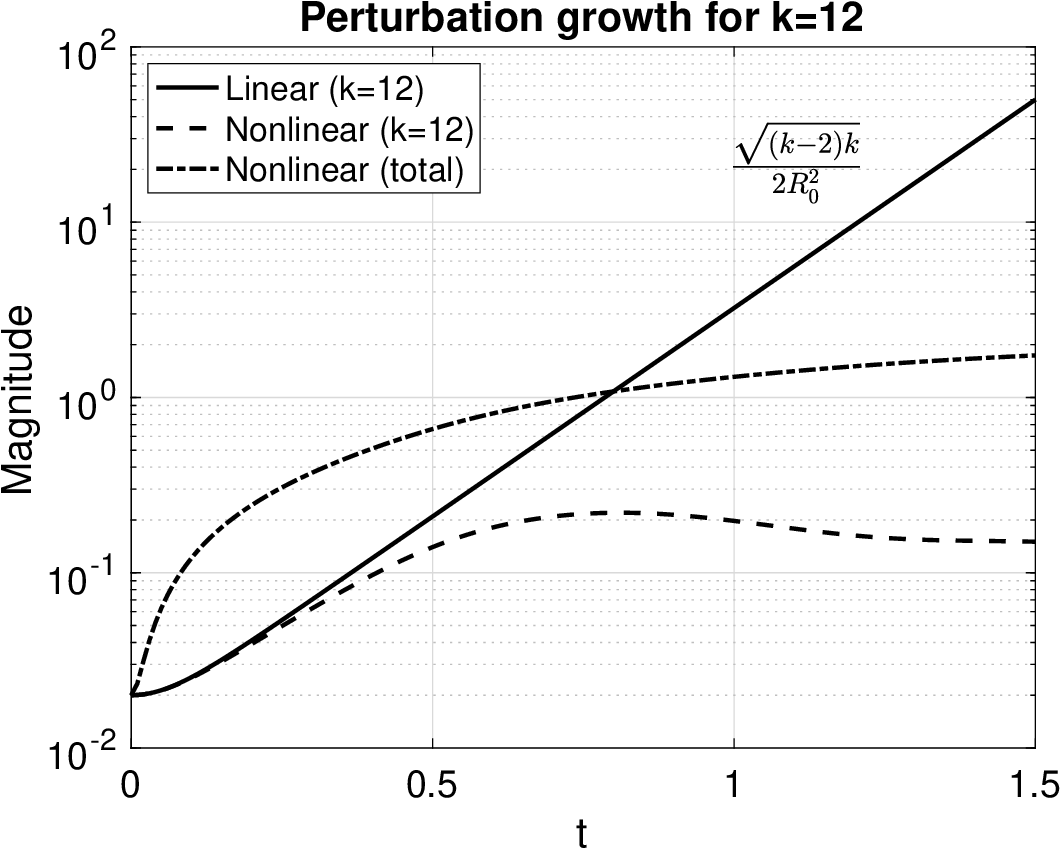}
        \caption{For initial condition with $k=12$.}\label{f:comp_k12}
    \end{subfigure}
    \caption{Solution components of $\xi$ on a semilog scale, demonstrating strong agreement between linear and nonlinear cases up to the time of singularity development.}
    \label{f:component_compare}
\end{figure}

Figure \ref{f:component_compare} shows the result on a semi-log scale for the $k=6$ and $k=12$ cases. In both cases, the linear solution displays the expected exponential growth like $e^{\sqrt{(k-2)k}/(2R_0^2)t}$, which appears linear on the semi-log scale. The time of singularity formation for the $k=6$ case is $t=1.25$; up to this time Figure \ref{f:comp_k6} shows strong agreement between the linear and nonlinear evolution of the 6th Fourier component. Similarly in the $k=12$ case, the linear and nonlinear evolution of the 6th Fourier component shows strong agreement up to singularity development at $t=0.43$. The agreement shown in Figure \ref{f:component_compare} is the key piece of evidence to argue that the linear dynamics are sufficient to the capture the initial wave-breaking mechanism of the Kelvin-Helmholtz instability. 

The total nonlinear perturbation magnitude grows much faster than the individual modes at short times because of coupling between Fourier modes. After the singularity develops, the linear solution continues to grow without bound, and diverges from the nonlinear simulation. This is due to the inability of the linear dynamics to capture spiral roll-up. After singularity development, the growth rate of individual $k=6$ and $k=12$ components, as well as the total nonlinear perturbation magnitude, begins to level off, approaching some maximum size of the spiral vortices. Although it takes longer to develop, the maximum vortex size in the $k=6$ case is slightly larger than in the $k=12$ case.

\section{Conclusion}

This article has provided a modern linear analysis for the stability of circular vortex sheets. We proved Theorem \ref{thm:main_instab}, which states that the full linearized evolution is unstable under small, smooth perturbations. We obtained the exponential growth rate of such a perturbation, and identified the short-wave intensified blowup which is the mathematical signature of the Kelvin-Helmholtz instability.

The growth rate obtained here agrees with that found in the classical works \cite{Moore1974, SaffmanVortex}, but the method is entirely different; instead of a potential flow argument, we used the Birkhoff-Rott equation as our starting point, ensuring that the correct flow conditions are applied. Our approach provides a completely explicit solution that lays a strong foundation for new results on existence, uniqueness, and singularity formation for rotational shear flows. To this point, we examined the linear evolution equation \eqref{e:full_lin_ev} to prove that a singularity can form in finite time from analytic initial data (Proposition \ref{prop:approx_sing}). Finally, we showed numerical evidence to suggest that the linear instability represents the wave-breaking mechanism of the fully nonlinear Kelvin-Helmholtz instability, up to the time of singularity formation.

Physically, we consider the circular vortex sheet to be a highly idealized model of a two-dimensional eddy. It is known that the the Kelvin-Helmholtz instability plays a role in the generation of such eddies from planar shear flows; we have shown strong evidence that the same mechanism can also catalyze a transition to turbulence via the rapid development of further instability from high-frequency geometric perturbations.

Our work provides a valuable tool for further study of rotational shear flows. Singularity formation of the nonlinear equation is thus far unproven for the circular sheet; the explicit nature of Equation \eqref{e:full_lin_ev} will aid in the construction of this proof. The effect of an expanding sheet is also an interesting question: can stretching of the sheet suppress the development of instability, as suggested by the form of blowup (and as discussed in \cite{Moore1974, SaffmanVortex}, although the form of their expanding vortex sheet cannot satisfy the Birkhoff-Rott equation)? It may also be possible to adapt the preceding analysis to study more complex and physically relevant vortex-type flows by building shear layers out of successive parallel vortex sheets with coupled dynamics.

\section*{Acknowledgements}

The authors would like to thank Ayman Said and Mohammad Farazmand for helpful discussions and valuable feedback. We also thank Volker Elling for an insightful comment which led to the discovery of an error in a previous version of this analysis. The first author is grateful to his current advisors, Gordon Zhang and Catherine Walker, for their continued support of this project. The second author acknowledges support from the Simons Foundation (MP-TSM).

\section*{Declaration of interests}
The authors report no conflict of interest.

\pagebreak

\printbibliography

\end{document}